\documentclass[twocolumn,showpacs,preprintnumbers,amsmath,amssymb]{revtex4}

\usepackage{graphicx}
\usepackage{dcolumn}
\usepackage{bm}

\begin{document}

\preprint{APS/123-QED}

\title{Meissner effect in a charged Bose gas with short-range repulsion}

\author{Shun-ichiro Koh}
 
\email{koh@kochi-u.ac.jp}

\affiliation{ Physics Division, Faculty of Education, Kochi University  \\
        Akebono-cho, 2-5-1, Kochi, 780, Japan 
}%

\date{\today}

\begin{abstract}
The question of whether the Bose-Einstein condensation (BEC) is a necessary condition of the  
Meissner effect is examined. The electromagnetic susceptibility of a 
charged Bose gas with short-range repulsion is studied using the perturbation theory 
with respect to the repulsive force.   With decreasing temperature, the 
Bose-statistical coherence grows, and prior to the BEC phase the  susceptibility  
shows a singularity implying the Meissner effect. This means that the BEC is a 
sufficient, but not a necessary, condition of the  Meissner effect in the 
charged Bose gas with short-range repulsion.  
\end{abstract}

\maketitle

\section{Introduction}
The Meissner-Ochsenfeld effect (which from now on we call the Meissner effect) is 
known as a striking phenomenon showing directly the existence of the Bose-Einstein 
condensate (BEC).  Actually, when cooling the metal in an applied 
magnetic field, just at the moment when the metal becomes 
superconducting, the metal begins to exclude the magnetic field.

A simple model exhibiting the Meissner effect is the charged Bose gas. 
Before the advent of the BCS model, Schafroth gave a clear definition of 
the Meissner effect using the charged Bose gas \cite{sch1}\cite{sch5}. 
The BCS model, starting from a realistic picture of the electrons in 
metal, established a realistic model exhibiting the Meissner effect \cite{bar}.

A common feature of these two models is an appearance of the 
off-diagonal long-range order (ODLRO) in low temperature such as \cite{yan}, 
\begin{equation}
	<\Phi(x)^{\dagger}\Phi(0)> \Longrightarrow  f^{\ast}(x)f(0), \qquad   |x|\rightarrow \infty.
	\label{е}
\end{equation}е
To make the problem clearer,  a model-independent derivation of the Meissner 
effect from fundamental principles and the ODLRO is desirable. 
Recently, such a derivation was attempted, with the 
result that the Meissner effect is derived from the gauge 
invariance and the ODLRO \cite{sew}\cite{nie}. This result means that the 
ODLRO is a sufficient condition of the  Meissner effect.

For the converse statement that the ODLRO is a necessary condition of the Meissner 
effect, however, the problem is not so simple.  Consider the following 
counterexample. In the two-dimensional systems, the order which one 
observes at zero temperature is not the ODLRO, but the 
off-diagonal  finite-range order (ODFRO), which falls off exponentially 
at a long distance \cite{hoh}. But its dynamical response is not an 
ordinary one. Experimentally, the thin-film superconductors 
exhibit the Meissner effect. Theoretically, the 
two-dimensional charged Bose gas,  although it does not show any 
mathematical singularity in the thermodynamic quantities, excludes the 
applied magnetic field at low temperature, which cannot be distinguished from 
the conventional Meissner effect in a practical sense \cite{may}. This example poses 
a question of whether the ODLRO is a necessary condition of the Meissner effect. In 
other words, we face another question of whether we must have a more 
general definition of the Meissner effect than our present one.

In this paper, we assert that the ODLRO is not a necessary condition of 
the Meissner effect. The Meissner effect occurs even at temperatures 
in which the many-body wave function obeying Bose statistics grows to a large 
but not yet macroscopic size (ODFRO) \cite{dia}. 

An objection to this assertion is that one does not observe in the 
superconductors the Meissner effect at $T>T_c$. The phenomenon  
occurring in the superconducting metals at the vicinity of $T_c$ is, however, 
 not a gradual growth of the coherent wave function  from a finite to a 
 macroscopic scale, but a formation of the Cooper pairs from two 
 electrons. Once the Cooper pair forms a composite boson at 
 low temperature and high density, the system does not change through the 
 pre-BEC state, but it immediately jumps to the BEC 
 phase \cite{flu}. Hence, the Meissner effect abruptly occurs at $T_c$ without any 
 precursor. In this meaning, the absence of the Meissner effect at 
 $T>T_c$ in the superconductors does not contradict the above assertion. 
 
  In principle, the charged Bose gas is a most elementary model of the 
 Meissner effect. But, we do not have any experimental example except for the 
 superconducting metal. This situation leads us to a preconceived 
 idea that the ODLRO is a necessary and sufficient condition of the 
 Meissner effect. Although the BEC is a most peculiar feature of Bose statistics, it 
does not automatically mean that all anomalous properties are attributed to the BEC. 
The bosons obey Bose statistics even when the Bose condensate is 
absent. Hence, to what extent these anomalous phenomena imply the 
existence of the Bose condensate is, logically, a problem to be considered separately.
From this standpoint, we can classify all phenomena in the dense cold bosons 
into two classes, the distinguishing factor being the necessity of the 
Bose condensate. The first is a class for which the Bose condensate is a 
necessary and sufficient condition, and the second is a class for which 
the Bose condensate is only a  sufficient one, thus being possible even 
when the Bose condensate is absent.  

 In real systems, the existence of the interaction between 
particles complicate the above classification.  An 
interplay between Bose statistics and the repulsive or attractive interaction 
determines the class of the phenomenon. 
 In this paper, we study the Meissner effect using the charged  
 Bose gas with short-range repulsion having the following hamiltonian,
 \begin{eqnarray}
&&H=\int d^3x\left(\frac{E^2+B^2}{2е}+|(\partial _{\mu}+ieA_{\mu})\Phi (x)|^2\right)  \nonumber\\
   && +g \sum_{p,p'}\sum_{q}\Phi_{p-q}^{\dagger}\Phi_{p'+q}^{\dagger}\Phi_{p'}\Phi_p , еее
	\label{е}
\end{eqnarray}
where $\Phi$ is the  spinless boson, and
 $g$ ($>0$) represents the short-range repulsive interaction due to the 
internal structure of the bosons ($\mu =x,y,z$). 

As will be explained in Sec.2A,  the Meissner effect occurs when the 
transverse excitation of the system is suppressed compared with the 
longitudinal excitation \cite{bay}. Following Feynman's argument 
\cite{fey}, Sec.2B explains that this suppression is a fundamental feature of  
Bose statistics,  being independent of the presence or absence of the 
BEC. In the charged ideal Bose gas ($g=0$), where the long-range Coulomb 
force is assumed to be screened by the opposite charges in the medium, 
the suppression of the transverse excitation comes only from Bose statistics, 
and it exhibits the Meissner effect only in the BEC phase. 
 In the charged Bose gas with short-range repulsion ($g>0$), however, the collective 
 excitation due to the repulsive interaction complicates the situation. As 
 a general property of the gas, the longitudinal excitation prevails over 
 the transverse one, thus destroying further the balance between 
the transverse and the longitudinal excitation. When this effect is 
added to the suppression due to Bose statistics, it is natural to ask whether the  
Meissner effect occurs before the many-body wave function grows to a 
macroscopic scale.

Sec.3 studies the Meissner effect by calculating the electromagnetic 
susceptibility of the charged Bose gas with short-range repulsion 
 using the perturbation expansion  with respect to the repulsive interaction $g$. 
To examine a meaning of the BEC for the Meissner effect, we cannot 
assume the Bose condensate from the beginning by replacing the operator 
of zero-momentum particle with a $c$-number \cite{bog}. 
Rather, it is crucially important to incorporate the 
Bose-statistical coherence of the many-body wave function without assuming 
the Bose condensate. Starting from the normal phase, the perturbation must be developed 
in such a way that as the  order of the expansion increases, the  
susceptibility includes a new effect due to a larger coherent wave function. 
 Sec.3 shows an occurrence of the  Meissner effect prior to the BEC phase 
 in cooling the charged Bose gas with short-range repulsion. Sec.4 
discusses the role of the repulsive interaction from a different point of 
view, and compare this Meissner effect with the conventional one in the superconductors.

\section{Meissner effect in the charged Bose gas} 
\subsection{Divergence of the susceptibility}

We apply to the charged Bose gas a weak stationary magnetic field derived 
from a transverse vector potential $A_{\mu}(q,\omega)e^{i\omega t}$.
 The current density within the linear response is, 
\begin{equation}
	\langle J_{\mu}(q,\omega)\rangle =-\frac{e^2}{m^2cе}[nm-\chi^T(q,\omega)]A_{\mu}(q,\omega)е,
	\label{е}
\end{equation}е
where $n$ and $m$ is  a number density and a mass of the particle, respectively. 
The first term in the right-hand side 
of Eq.(3) is the usual diamagnetic current, and the second one is the paramagnetic 
term arising from a change in the system's wave function induced by 
$A_{\mu}(q,\omega)$. The second term $\chi^T(q,\omega)$ is a transverse part of 
the Fourier component of the current-current response tensor, 
\begin{equation}
	\chi_{\mu\nu}(q,\omega _n)=\frac{1}{Vе}\int_{0}^{\betaе}d\tau \exp(i\omega_n\tau)
	                          \langle T_{\tau}J_{\mu}(q,\tau)J_{\nu}(q,0)\rangleеее,
	\label{е}
\end{equation}е
where
\begin{equation}
	J_{\mu}(q,\tau)=\sum_{p,n} 
	\left(p+\frac{q}{2е}\right)_{\mu}\Phi_p^{\dagger}\Phi_{p+q}e^{i\omega _n\tau}еее,
	\label{е}
\end{equation}е
($\hbar =1$), and defined as,
\begin{equation}
	\chi_{\mu\nu}(q,\omega )=\frac{q_{\mu}q_{\nu}}{q^2е}\chi^L(q,\omega)     
	                    +\left(\delta_{\mu\nu}-\frac{q_{\mu}q_{\nu}}{q^2е}\right)е\chi^T(q,\omega) .
	\label{е}
\end{equation}е
Inside of the matter, Eq.(3) is added to the Maxwell equation in vacuum,
\begin{equation}
	-\left(\frac{\omega 
	^2}{c^2е}-q^2е\right)A_{\mu}^{ex}(q,\omega)=\frac{4\pi}{cе}\langle 
             	J_{\mu}^{(0)}(q,\omega)\rangle ее,
	\label{е}
\end{equation}е
as an additional source term. Hence, one obtains a following relationship 
between $A_{\mu}(q,\omega)$ in matter and  $A_{\mu}^{ex}(q,\omega)$ in vacuum, 
\begin{equation}
	A_{\mu}(q,\omega)=\frac{A_{\mu}^{ex}(q,\omega)}
	                  {1-\displaystyle{\frac{4\pi e^2}{m^2c^2е}
	                                 \left(\frac{nm-\chi^T(q,\omega)}{\omega^2/c^2-q^2е}\right)ееее}}е.
	\label{е}
\end{equation}е
For the static limit, from gauge invariance, one knows $nm=\chi^L(q,0)$ 
relating the longitudinal current-current response tensor 
$\chi^L(q,\omega)$ with number density $n$ and mass $m$ of the particle 
\cite{noz}. Hence, one gets,  
\begin{equation}
	A_{\mu}(q,0)=\frac{A_{\mu}^{ex}(q,0)}
	                  {1+\displaystyle{\frac{4\pi e^2}{m^2c^2е}
	                  \left(\frac{\chi^L(q,0)-\chi^T(q,0)}{q^2е}\right)ееее}}е.
	\label{е}
\end{equation}е
Equation.(9) means that, when the balance between the transverse and the 
longitudinal excitation is destroyed in the static and uniform limit  
($\lim _{q\rightarrow 0}[\chi^L(q,0)-\chi^T(q,0)]\ne 0$), the denominator 
in the right-hand side of Eq.(9) diverges as $q\rightarrow 0$, thus 
vanishing $A_{\mu}(q,0)$ and $H=\nabla \times A$ at $q\rightarrow 0$ 
in matter (Schafroth's criterion of the Meissner effect).

 From now on, we express a  term proportional to $q_{\mu}q_{\nu}$ in 
 $\chi_{\mu\nu}$ by $\hat{\chi}_{\mu\nu}$ such as,
 \begin{eqnarray}
&&	\chi_{\mu\nu}(q,\omega)=\delta_{\mu\nu}\chi^T(q,\omega)
	                  +q_{\mu}q_{\nu}\left(\frac{\chi^L(q,\omega)-\chi^T(q,\omega)}{q^2е}\right)е\nonumber\\
	                &&  \equiv \delta_{\mu\nu}\chi^T+\hat{\chi}_{\mu\nu}. 
	\label{е}
\end{eqnarray}
 Hence, Eq.(9) is expressed as, 
\begin{equation}
	A_{\mu}(q,0)=\frac{A_{\mu}^{ex}(q,0)}
	                  {1+\displaystyle{\frac{4\pi e^2}{m^2c^2е}
	                  \left(\frac{\hat{\chi}_{\mu\nu}(q,0)}{q_{\mu}q_{\nu}е}\right)ееее}}е.
	\label{е}
\end{equation}е

(1) In the charged ideal Bose gas,  $\hat{\chi}_{\mu\nu}$ is given by, 
\begin{equation}
	\hat{\chi}_{\mu\nu}(q,\omega)=-\frac{q_{\mu}q_{\nu}}{4е}е
	                          \sum_{p}\frac{f(\epsilon (p))-f(\epsilon (p+q))}
	                                       {\omega+\epsilon (p)-\epsilon (p+q)е}е,
\end{equation}е 
where $\epsilon (p)$ is the kinetic energy $p^2/2m$ of the spinless boson, and 
$f(\epsilon (p))$ is the Bose-Einstein distribution.
In the BEC phase (chemical potential $\mu=0$), $f(\epsilon (p))$ in Eq.(12) is a macroscopic 
number for $p=0$ and nearly zero for $p\ne 0$. Thus, in the sum over $p$ in the 
right-hand side of Eq.(12), only two terms corresponding to $p=0$ and 
$p=-q$ remain, with a result that, 
\begin{equation}
	\hat{\chi}_{\mu\nu}(q,0)=mn_0е\frac{q_{\mu}q_{\nu}}{q^2е}е,
\end{equation}е 
where $n_0$ is the number density of the $p=0$ Bose particles. 
In view of $q^{-2}$ in Eq.(13), the charged ideal Bose gas in the BEC 
phase exhibits the Meissner effect. In the normal phase ($\mu < 0$), 
however, when the sum over $p$ in Eq.(12) is carried out by replacing it 
with an integral, one notices that $q^{-2}$ dependence disappears in the result. 
Hence, in the charged ideal Bose gas, the BEC is a necessary and 
sufficient condition of the Meissner effect.
In this case, the penetration depth $\lambda _0(T)$, defined in 
Eq.(3) by $J_{\mu}(q,0)=-(c/4\pi)(1/\lambda _0^2(T)) A_{\mu}(q,0)$ at $q\rightarrow 0$, 
is $\lambda _0(T)=n_0(T)^{-0.5}\sqrt{mc^2/4\pi e^2}$. 

(2) To analyze the Meissner effect in the charged Bose gas with 
short-range repulsion, instead of
$\lim _{q\rightarrow 0}[\chi^L(q,0)-\chi^T(q,0)]\ne 0$,
we regard a divergence of the coefficient of $q_{\mu}q_{\nu}$ in 
$\chi_{\mu\nu}(q,0)$ at $q\rightarrow 0$ (Eq.(10)), as a general 
definition of the Meissner effect.

\subsection{Effects of Bose statistics}
The physical explanation of $\lim _{q\rightarrow 
0}[\chi^L(q,0)-\chi^T(q,0)]\ne 0$ in the Bose gas dates back to  
Feynman's argument on the scarcity of the excitation in the liquid helium 
4 \cite{fey}. We recapitulate his explanation of how Bose statistics affects the 
many-body wave function in configuration space. 

\begin{figure}
\includegraphics [scale=0.5]{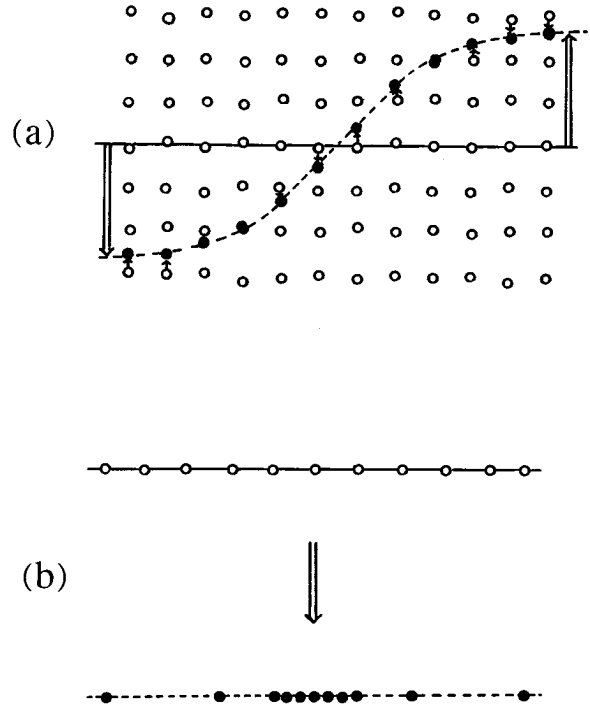}
\caption{\label{fig:epsart}
 A schematic picture of (a) transverse displacement, and (b) 
 longitudinal displacement in coordinate space. White circles represent 
 the bosons in the initial configuration.}
\end{figure}

(1) When the gas is in the BEC phase ($\mu =0$), the wave function has 
the permutation symmetry  on a macroscopic scale. Consider a collective 
transverse displacement moving particles on a solid straight line in 
Fig.1(a) to that on a dotted wavy curve (black circles). 
At first, it seems a configuration change on a large scale. This 
displacement is, however, reproduced by a set of slight displacements 
(depicted by short arrows) moving the particle   
close to the wavy curve in the initial configuration to the exact 
positions on the curve. The transverse displacement like Fig.1(a) 
changes  only slightly  the local density of particles (defined in some 
volume). Hence, at any part of the wavy curve, it is possible 
to find in the initial configuration a particle close to the wavy curve. 
 In Bose statistics, due to permutation symmetry, one cannot 
distinguish between one particle on the wavy curve  which comes 
from the position close to the wavy curve by the short arrow  
and another particle which comes from the straight line  by the long arrow. 
Even if  the displacement from the straight to the wavy line is a large 
displacement in the classical statistics, it is only a slight 
displacement in Bose statistics (short arrows). Hence, the 
excited state lies in a small distance from the ground state in 
configuration space. Since the excited state is orthogonal to the ground 
state, the wave function corresponding to the excited state must 
spatially oscillate. Accordingly, the many-body wave 
function of the transversely excited state oscillates within a small 
distance in configuration space. Since the kinetic energy of the system 
is determined by the gradient of the wave function, this means that the 
energy of the transverse excitation is not small even at $q=0$, leading 
to the scarcity of the low-energy transverse-excitation density. This is a reason for 
$\chi^T(q,0)\rightarrow 0$ as $q\rightarrow 0$ at low temperature.  

On the other hand, for the longitudinal displacement like Fig.1(b), we 
see a different situation. Since the longitudinal displacement changes 
the local density so much, it is not possible to always find in the initial 
configuration a particle which is close to a given particle after displacement. 
(Even if it is possible for only a few particles after displacement, it is 
not possible for all particles.) Hence, one can not reproduce the configuration 
pattern after displacement only by permutation of the initial pattern 
(a lower pattern in Fig.1(b) can not be reproduced with an upper 
pattern). This means that the longitudinally-displaced configuration is 
far from the initial one in configuration space. Hence, the wave function of the 
longitudinally excited state gradually oscillates in a long distance. 
Applying the definition of the kinetic energy to this case, one concludes that  
the low-energy longitudinal excitation is possible.  This is a reason for 
$\chi^L(q,0)\ne 0$ at $q\rightarrow 0$. The above explanation asserts 
that Bose statistics is an essential reason for the Meissner effect.  

(2)  When the gas is at high temperature ($\mu \ll 0$),  the wave 
function has the permutation symmetry only in the limited area. 
Accordingly, in the transverse 
displacement like Fig.1(a), one can not regard the wave function  
after permutation as an equivalent of the initial wave function.  
The displacement which was regarded as a small displacement in the BEC 
phase becomes a large one in the normal phase. Since the difference 
between the transverse and the longitudinal excitation discussed in 
(1) vanishes, the Meissner effect disappears at high temperature.

(3) From our viewpoint, a remarkable state of the Bose gas lies at the 
vicinity of the BEC transition temperature in the normal phase ($\mu \leq 
0$) \cite{koh}. The coherent many-body wave function grows to a large size but not 
yet to a macroscopic one. In such a situation, whereas the mechanism in 
(2) works for the large displacement extending over two different wave 
functions, the mechanism in (1) works for the small displacement within 
a single wave function. 

As shown in Eq.(13), the BEC is the necessary and sufficient condition  
of the Meissner effect for the charged ideal Bose gas. This conclusion is, 
however, applicable only to a simplified model 
like the charged ideal Bose gas. An inclusion of the repulsive 
interaction will change the situation, which will be studied in Sec.3.

\section{Formalism}

To examine the normal Bose gas at the vicinity of the BEC transition 
temperature, we can not use the conventional method of 
 assuming the Bose condensate by replacing the operator of zero 
 momentum particle with a $c$-number from the beginning.
Consider the current-current response tensor of the charged Bose gas at the 
vicinity of the BEC transition temperature as illustrated in Fig.2. The $\chi_{\mu\nu}$ in 
coordinate space includes the following form,
 \begin{eqnarray}
	&&\langle G|T_{\tau}J_{\mu}(x,\tau)J_{\nu}(0,0)|G\rangle   \nonumber\\
      &&= \frac{\langle 0|T_{\tau}\hat{J}_{\mu}(x,\tau)\hat{J}_{\nu}(0,0)
 	              exp\left[-\displaystyle {\int_{0}^{\betaе}}d\tau \hat{H}_I(\tau)е\right]|0\rangleе}
 	        {\langle 0|exp\left[-\displaystyle {\int_{0}^{\betaе}}d\tau  \hat{H}_I(\tau)е\right]|0\rangleе},
	\label{е}
\end{eqnarray}
 where $|G\rangle $ is a ground state of $\sum_{p}\epsilon (p)\Phi_p^{\dagger}\Phi_p
 +g \sum_{p,p'}\sum_{q}\Phi_{p-q}^{\dagger}\Phi_{p'+q}^{\dagger}\Phi_{p'}\Phi_p$. 
 The excitation energy due to the long-range Coulomb force is too high to 
 affect the dynamics of the bosons in the Meissner effect \cite{cou}. 
 Hence, the photon-boson vertex appears only at an initial and a final 
 vertex of the $\chi_{\mu\nu}$, and $H_I(\tau)$ in Eq.(14) represents 
 $g\sum_{p,p'}\sum_{q}\hat{\Phi}_{p-q}^{\dagger}\hat{\Phi}_{p'+q}^{\dagger}\hat{\Phi}_{p'}\hat{\Phi}_p$    
 (dotted lines in Fig.2).   
 In the ground state of Bose gas, due to the repulsive interaction $g$ 
 contained in $exp(-i\smallint  H_I(\tau)d\tau)$ of Eq.(14), 
 one-particle excitations frequently occur  as illustrated by an upper 
 bubble with a dotted line in Fig.2(a). The current-current response 
 tensor of such a medium is depicted by a lower 
 bubble in Fig.2(a). (The black and the white small circle in Fig.2 
 represents the vector and the scalar vertex respectively.) 

\begin{figure}
\includegraphics [scale=0.3]{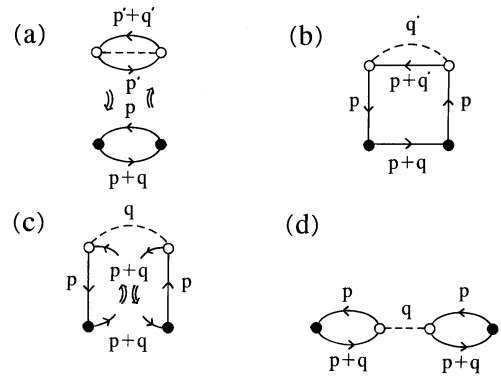}
\caption{\label{fig:epsart}
 The first-order Feynman diagram of the current-current 
           response tensor among the one-particle 
           excitations of the repulsive Bose gas in Eq.(14). (The solid and the 
           dotted lines represent the boson and the repulsive interaction, 
           respectively.) Particle exchange between the tensor and the 
           medium by Bose symmetry is explained in four steps from (a) to (d).}
\end{figure}

When $|G>$ is the ground state at the vicinity of the BEC transition 
temperature in the normal phase, the perturbation must be developed 
in such a way that as the  order of the expansion increases, the  
susceptibility gradually includes a new effect due to Bose statistics.
  An important feature of the response  to the photon by the large coherent many-body wave 
  function is that bosons appearing in the response tensor 
 also participate in the coherent wave function of the medium. 
(The response tensor and the medium form a coherent wave function as a whole.) 
 Hence, we must seriously  consider the influence of Bose statistics on 
the graph like Fig.2(a). When one of the two bosons in the lower bubble 
and one of the two bosons in the upper bubble have a same momentum 
($p=p'$), and another bosons in the lower and the upper bubble have 
another same momentum ($p+q=p'+q'$), a graph made by 
exchanging these two particles must be included in the expansion of 
$\chi_{\mu\nu}$. Such a graph is obtained in two steps as illustrated in Fig.2. 
When two boson lines with $p$ and $p'$($=p$) is exchanged in Fig.2(a), a 
square appears in Fig.2(b). Further, when two boson lines with 
$p+q$ and $p+q'$($=p+q$) is exchanged in Fig.2(c), a graph linked by the 
repulsive interaction appears in Fig.2(d). The resulting  
current-current correlation tensor has a form elongated by the density-density 
correlation.  

A contribution of Fig.2(d) to $\chi_{\mu\nu}$ is given by,
\begin{eqnarray}
	&&\chi_{\mu\nu}^{(1)}(q,\omega)=g\sum_{p}(p+\frac{q}{2е}е)_{\mu}(p+\frac{q}{2е}е)_{\nu}  \nonumber\\
	            &&\times    \left[-\frac{f(\epsilon (p))-f(\epsilon (p+q))}
	                            {\omega+\epsilon (p)-\epsilon (p+q)е} \right]^2еее.
	\label{е}
\end{eqnarray}
With decreasing temperature, the coherent wave function grows to a large 
size, and the particle exchange due to Bose statistics like Fig.2 occurs 
many times. Hence, one can not ignore the higher-order term $\chi_{\mu\nu}^{(n)}$ 
corresponding to the larger coherent wave function. Particularly 
important is the coherent wave function consisting of $p=0$ bosons. 
Correspondingly, a quantity of most interest  which contribute to 
$\hat{\chi}_{\mu\nu}$ in Eq.(10) is a term proportional to $q_{\mu}q_{\nu}$ in 
$\chi_{\mu\nu}^{(n)}(q,0)$, the $p=0$ component of which has the following form; 
\begin{equation}
	\hat{\chi}_{\mu\nu}^{(n)}(q,0)=\frac{q_{\mu}q_{\nu}}{2е}еg^nF_{\beta}(q)^{n+1},
	\label{е}
\end{equation}е
where
\begin{widetext}
\begin{equation}
	F_{\beta}(q)=\exp(\beta\mu)
	                  \left( 
	                  \frac{(1-\exp(\beta\mu))^{-1}-(\exp(\beta\epsilon (q))-\exp(\beta\mu))^{-1}}
	                         {\epsilon (q)ее}  \right)е.
	\label{е}
\end{equation}е
\end{widetext}
$F_{\beta}(q)$ is a positive monotonously decreasing function of $q^2$ which approaches zero as 
$q^2\rightarrow \infty$. As $\mu\rightarrow 0$, $|F_{\beta}(q)|$ increases at any $q$.

The general definition at the end of Sec.2A says that, when a power series in $g$ 
\begin{equation}
	\hat{\chi}_{\mu\nu}(q,0)=\frac{q_{\mu}q_{\nu}}{2е}е\sum_{n=0}^{\inftyе}g^nF_{\beta}(q)^{n+1}
	                       \equiv\frac{q_{\mu}q_{\nu}}{2е}\sum_{n=0}^{\inftyе}a_ng^n,
	\label{е}
\end{equation}е
 diverges at $q\rightarrow 0$, the Meissner effect occurs. At high temperature in 
normal phase ($\beta\mu \ll 0$), a small $F_{\beta}(q)$ guarantees the 
convergence of $\hat{\chi}_{\mu\nu}(q,0)$ (Eq.(18)). With 
decreasing temperature ($\mu\rightarrow 0$), however, a gradual increase  
of  $F_{\beta}(q)$ makes the higher-order term significant, finally 
leading to the divergence  of  $\hat{\chi}_{\mu\nu}(q,0)$. 
A convergence radius $r_c$ of the power series  
$\sum_{n=1}^{\infty}a_nx^n$ is given by
\begin{equation}
	\frac{1}{r_cе}е=\lim _{n\to \infty}|a_n|^{1/n},
	\label{е}
\end{equation}е
(Cauchy-Hadamard's theorem). Applying this theorem to Eq.(18) 
($a_n=F_{\beta}(q)^{n+1}$), one obtains $1/r_c=F_{\beta}(q)$. Hence, as a 
convergence condition: $0<g<r_c$, we have 
\begin{equation}
	|gF_{\beta}(q)|<1.
	\label{е}
\end{equation}е
Generally, for the Meissner effect to occur, the balance between the longitudinal 
and the transverse excitation must be destroyed on a macroscopic scale 
($q\rightarrow 0$). Together with the Bose-statistical coherence, the 
repulsive interaction enhances this tendency, finally violating the 
condition of Eq.(20).

 An expansion form of $F_{\beta}(q)$ around $q^2=0$  is given by
\begin{eqnarray}
&&	F_{\beta}(q)=\frac{\beta\exp(\beta\mu)}{(1-\exp(\beta\mu))^2е} \nonumber\\
	            &&   -\frac{\beta ^2}{2е}
	                \exp(\beta\mu)\frac{(1+\exp(\beta\mu))}{(1-\exp(\beta\mu))^3е}\epsilon(q)+\cdots.е
	\label{е}
\end{eqnarray}
At high temperature, Eq.(20) is satisfied, and $\hat{\chi}_{\mu\nu}(q,0)$ has the 
following form, 
\begin{equation}
	\hat{\chi}_{\mu\nu}(q,0)=\frac{q_{\mu}q_{\nu}}{2е}е
	                 \frac{F_{\beta}(q)}{1-gF_{\beta}(q)е}.ее
	\label{е}
\end{equation}е
In cooling the gas, $|F_{\beta}(q)|$ increases monotonously, and the 
convergence condition is first violated at $q=0$ when 	
$|gF_{\beta}(0)|=1$, that is, using Eq.(21), 
\begin{equation}
     g\beta=4\sinh ^2\left(\frac{\beta\mu}{2е}е\right)е .
	\label{е}
\end{equation}е
 From now, we call $T_0$ an onset temperature of the 
Meissner effect satisfying Eq.(23).  In view of Eq.(23), we conclude that 
the charged Bose gas with short-range repulsion ($g>0$) shows the Meissner effect prior to 
the BEC phase ($\mu <0$), thus leading to $T_{BEC}<T_0$. At $T=T_0$, substituting 
 Eq.(21) in Eq.(22) with Eq.(23), we get
\begin{equation}
	\hat{\chi}_{\mu\nu}(q,0)=\frac{m}{\sinh |\beta\mu |е}е
	                              \frac{q_{\mu}q_{\nu}}{q^2е}е,
\end{equation}е 
which shows the Meissner effect in Eq.(11).

Compared with Eq.(13), the number density $n_0$ of $p=0$ boson which is 
a macroscopic number at $T_{BEC}$ is replaced by $0.5/\sinh |\beta\mu |$ 
in Eq.(24). Accordingly, the penetration depth has a form such as 
$\lambda (T)=\sqrt{\sinh |\beta\mu |}\sqrt{mc^2/4\pi e^2}$.  Since
 $\mu <0$ at $T_0$, $\lambda (T_0)$ is larger than  
 $\lambda _0(T_{BEC})$ in the ideal Bose gas, because at $T_0$ only $p=0$ 
 boson whose number is large, but 
 not yet macroscopic is responsible for this Meissner effect. Hence, 
 at the onset temperature $T_0$, the magnetic field penetrates deeply 
 into the gas. (It is in contrast with the abrupt exclusion of the 
 magnetic field, that is, a very small $\lambda (T_c)$ in the superconductor.)

(a) In the weak-coupling case $(g\simeq 0)$, we can make a rough estimation 
of the condition of Eq.(23). Using $\mu _0(T)$ 
of the ideal Bose gas for $\mu (T)$ in Eq.(23) such as
\begin{equation}
	\beta\mu _0(T)=-\left(\frac{2.61}{2\sqrt {\pi}е}е\right)^2\frac{T_{BEC}}{Tе}
	                  \left[\left(\frac{T}{T_{BEC}е}\right)^{1.5}е-1\right]^2еее,
	\label{е}
\end{equation}е
 we can determine $T_0$ as a solution of 
 simultaneous equations (23) and (25).  Figure.3 shows a phase diagram 
 in which the onset temperature $T_0$ of the Meissner effect is plotted at a given 
 strength of the repulsive interaction $g$. (Since only the relative 
 strength of $g$ to the particle mass has a meaning, $g/k_BT_{BEC}$ is plotted.)
  Figure.3 confirms that for the charged ideal Bose gas ($g=0$), the 
  Meissner effect occurs just at $T=T_{BEC}$, but it shows that the charged 
  Bose gas with short-range repulsion ($g \ne 0$) begins to exclude the applied 
  magnetic field prior to the BEC phase in cooling. (Although the 
  repulsive interaction suppresses the $T_{BEC}$, it does 
 not change the overall feature of Fig.3.) The BEC is not a necessary 
 condition of the Meissner effect for the charged Bose gas with 
 short-range repulsion.

\begin{figure}
\includegraphics [scale=0.3]{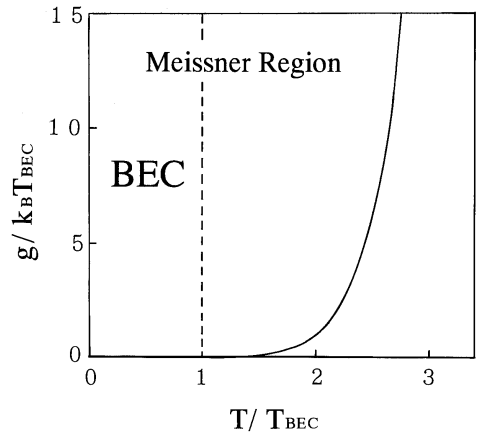}
\caption{\label{fig:epsart}
A phase diagram showing the onset temperature of the 
  Meissner effect $T_0$ at a given strength of the repulsive interaction $g$. Both 
  quantities are normalized by $T_{BEC}$ and $k_BT_{BEC}$, respectively.}
\end{figure}

(b) At high temperature ($T/T_{BEC} \gg 1$), the repulsive interaction 
which is necessary for the Meissner effect to occur in Eq.(23) increases as 
$T\exp [a(T/T_{BEC})^2]$. In view of the exponential factor, the Meissner 
effect does not occur at  high temperature in real materials, thus not 
affecting the essential feature of the Meissner effect, a phenomenon in 
low temperature.
 
(c) When cooling the system further below $T_0$ in Fig.3 at a given $g$ , 
  the $\hat{\chi}_{\mu\nu}(q,0)$ (Eq.(18)) diverges not only at
 $q=0$, but also for a region $0<q\leq q_c(T)$, where $q_c(T)$ is 
 determined by a condition $|gF_{\beta}(q_c(T))|=1$ at a given $T$.
  For $q_c(T)<q$, the $\hat{\chi}_{\mu\nu}(q,0)$ converges even 
  below $T_0$. To obtain the overall feature of 
  $\hat{\chi}_{\mu\nu}(q,0)$ below $T_0$,  we get a concrete form of 
  $\hat{\chi}_{\mu\nu}$ by expanding 
 $1-gF_{\beta}(q)$ in the  right-hand side of Eq.(22) with respect to 
 $q^2-q_c(T)^2$. Hence one gets the following result:
 \begin{eqnarray}
&& \hat{\chi}_{\mu\nu}(q,0) \nonumber\\
      &&=\left\{
                \begin{array}{ll} 
                  \displaystyle {\infty}, & ее
                    \quad 0\leq  q\leq q_c(T), \\ [0,2cm]
                 \frac{\displaystyle{-q_{\mu}q_{\nu}}}{\displaystyle{2g\left(\frac{\partial ln 
                 F_{\beta}(q)}{\partial q^2е}е\right)_{q_c}ее}
                         (q^2-q_c(T)^2)ее}, & ее
                     \quad  q_c(T)<q . 
                 \end {array}  \right.    	
      \label{е}
\end{eqnarray}
 Equation.(26) in Eq.(11) means that the charged Bose gas with short-range repulsion
excludes not only the uniform applied magnetic field $(q=0)$, but also the 
spatially modulated magnetic field $H(q)$ with $0< q\leq q_c(T)$. 
 
 With decreasing temperature, the number of particles with $p\ne 0$ which 
 satisfies the divergence condition of $\hat{\chi}_{\mu\nu}(q,0)$ 
 gradually increases. Hence such particles participate in excluding the 
 applied magnetic field, leading to a gradual decrease of the penetration depth.

The $q_c(T)$ reminds us of the coherence length $\xi$ in the superconductor, because 
$q_c(T)$ and $\xi^{-1}$ have a similar property: As they become 
larger, the Bose-statistical coherence grows. (For more comparison, see 
Sec.4.) Here, we discuss the temperature dependence of $q_c(T)$.

(1) Just below $T_0$,  $q_c(T)$ in Eq.(26) is small, and we can obtain 
the temperature-dependence of $q_c(T)$ as follows. 
We expand $F_{\beta}(q)$ in the condition of $|gF_{\beta}(q)|=1$ 
 with a small $q_c(T)$ as in Eq.(21). Taking only the first and the 
 second term in the right-hand side of Eq.(21), we get an approximate form of $q_c(T)$ as
\begin{equation}
   q_c(T)^2=4mk_BT\tanh\frac{|\beta\mu(T)|}{2е}
                 \left[1-\frac{4k_BT}{gе}\sinh ^2\left(\frac{\beta\mu}{2е}е\right)ее\right]еее	.
	\label{е}
\end{equation}е
At $T=T_0$, $q_c(T)$ is zero because of Eq.(23). At $T<T_0$, 
$q_c(T)$ is expanded with respect to $T_0-T$.  In the case of weak 
repulsive interaction, using $\mu _0(T)$ (Eq.(25)) as $\mu (T)$, one gets
\begin{equation}
   q_c(T)=2\sqrt{Amk_BT\tanh\frac{|\beta\mu_0(T)|}{2е}}
                 \sqrt{1-\frac{T}{T_0е}е}еее,	
	\label{е}
\end{equation}е        
where
\begin{equation}
	A=1+1.08(\alpha^2-0.5\alpha^{0.5}
	            -0.5\alpha^{-1})\sqrt{\frac{4k_BT_0}{gе}+1е}е,
	\label{е}
\end{equation}е
and  $\alpha=T_0/T_{BEC}$.

(2) When the decreasing temperature reaches $T=T_{BEC}$, $F_{\beta}(q)$ 
(Eq.(17)) diverges at all $q$, thus leading 
to a divergence of $\hat{\chi}_{\mu\nu}(q,0)$ at  all $q$. Hence 
$q_c(T)$ in Eq.(26)  goes to $\infty$ in the BEC phase.

\section{Discussion}
(1) In this paper, we explained the mechanism in which the repulsive 
interaction destroys the balance between  the longitudinal and the transverse 
excitation on a macroscopic scale. This explanation can be viewed from a 
different point of view.

Generally, the influence of the repulsive interaction on the dynamical 
properties of the boson systems differs from that on the thermal 
properties.  For the thermal properties, the repulsive interaction 
generally suppresses the growth of the Bose-statistical coherence. For 
example, it is known that due to the strong short-range repulsion, not 
all helium 4 atoms participate in the BEC even at zero temperature 
(depletion effect).

 For the dynamical properties such as the superfluidity, however, the repulsive 
interaction generally enhances singular properties due to Bose 
statistics. For the ideal Bose gas, the particle with non-zero  momentum 
behaves like a free particle (except for the statistical constraint) 
to the external dynamical perturbation.
For the  repulsive Bose gas, however, the particles are likely to spread 
uniformly in coordinate space due to the repulsive force. This feature 
makes the particles with $p\ne 0 $ behave similarly with other particles, 
especially with the particle having zero momentum.
 If it behaves differently from others, a resulting 
locally high density of particle raises the interaction energy.  This is a 
reason why all particles seem to show singular dynamical behavior
even at temperature in which not all particles participate in the BEC.
(This fact corresponds to the well-known property that the one-particle 
energy spectrum changes from $p^2/2m$ to $sp$ due to the repulsive interaction, 
enhancing the stability of the superfluid flow to the thermal 
disturbance by raising the energy barrier around $p=0$.) 

The Meissner effect is a dynamical response of the Bose gas to the 
external perturbation (the applied static magnetic field) as well. 
The boson excited from the $p=0$ to the $p\ne 0 $ state does not 
respond as a free particle to the magnetic field, but it excludes the 
magnetic field cooperatively with the $p=0$ bosons by the repulsive 
interaction. Hence, even when the Bose-statistical coherence 
grows to a large but not to a macroscopic scale (ODFRO), the charged  
Bose gas with short-range repulsion shows the  Meissner effect. In this 
meaning, as well as the superfluidity, the  Meissner effect is another 
example of the mechanism in which the repulsive interaction enhances the 
singular dynamical response of the boson system. 

(2) As discussed in Sec,1, the Meissner effect occurring in the charged 
Bose gas with short-range repulsion differs from that in the superconductors in some 
respects. Comparing two different Meissner effects is useful for a deeper 
understanding of them. 

Let us replace the boson operator $\Phi_p$ in the right-hand side of Eq.(5) with 
the fermion operator $\Psi _{p,\sigma}$, and use such a $J_{\mu}(q,\tau)$ in Eq.(4). By 
factorizing the resulting $\langle T_{\tau}J_{\mu}(q,\tau)J_{\nu}(q,0)\rangle$ into a 
product of the two anomalous fermion Green's functions, we get
\begin{eqnarray}
&&	\chi_{\mu\nu}(q,0)=\frac{1}{2е}е\intе\frac{d^3p}{(2\pi)^3е}е
	                     (p+\frac{q}{2е}е)_{\mu}(p+\frac{q}{2е}е)_{\nu} \nonumber\\
	               && \times   \left(1-\frac{\epsilon _p\epsilon _{p+q}+|\Delta|^2}{E_pE_{p+q}е}е\right)е
	                     \frac{1}{E_p+E_{p+q}е}ее,
	\label{е}
\end{eqnarray}
where $\epsilon _p=p^2/2m-\mu $, $\Delta$ is the order parameter, and 
$E_p=\sqrt{\epsilon _p^2+\Delta ^2}$.  
Performing the integral over $p$ in Eq.(30) yields
\begin{equation}
	\chi ^T(q,0)=(nm-a|\Delta|^2)-bq^2,
	\label{е}
\end{equation}е
where $nm=\chi^L(q,0)$, and $a$ and $b$ is a parameter independent of $q^2$.
Substituting Eq.(31) in the definition of Eq.(10), one obtains
\begin{equation}
	\hat{\chi}_{\mu\nu}(q,0)=q_{\mu}q_{\nu}\left(\frac{a|\Delta|^2}{q^2е}+bе\right)ее,
\end{equation}е 
which means that only below $T_c$ ($|\Delta| \ne 0$), 
$\hat{\chi}_{\mu\nu}(q,0)$ shows the mathematical singularity like 
$q^{-2}$, leading to the Meissner effect. 

As well as the charged ideal Bose gas, the BCS model does not show the 
mathematical singularity in $\hat{\chi}_{\mu\nu}(q,0)$ prior to the BEC, 
but after the BEC, only the first-order term in the perturbation expansion of 
$\chi_{\mu\nu}(q,0)$ is enough to show the $q^{-2}$ singularity. In the 
BCS model, the particle interaction is assumed to be completely 
diagonalized by the formation of the Cooper pair, thus leaving no 
residual interaction between the pairs. This fact makes the next-order 
calculation of the susceptibility unnecessary. On the other hand, the 
charged Bose gas with short-range repulsion shows another type of 
singularity prior to the BEC in the normal phase, if the  perturbation expansion of 
$\chi_{\mu\nu}(q,0)$ is summed to infinite  order.

In the superconductors, two different kinds of the coherence length are 
known. The first is $\xi _0$ appearing in the Pippard nonlocal 
electrodynamics, and the second is $\xi (T)$ appearing in the 
Ginzburg-Landau equation. They are not the same quantities, but both  
quantities represent the smallest size of the wave packets which the supercurrent 
carriers can form, thus reflecting an extent of transformation of 
individual particles from fermion to boson. Due to the composite-boson 
nature underlying the Cooper pairs, the type-2 superconductors show the 
complicated nonuniform response to the uniform applied magnetic field.  

On the other hand, the response of the charged Bose gas with short-range 
repulsion to the magnetic field is different, because it consists of the 
elementary boson. ($\xi _0$ or $\xi (T)$ corresponds to the particle 
radius in this case.) The critical wave number $q_c(T)$ defined as 
Eq.(28) is an inverse of another type of the  coherence length. The 
$1/q_c(T)$ represents a characteristic length of the 
spatial variation of magnetic field. Generally, compared with the uniform 
field case, screening of the spatially modulated applied magnetic field needs the 
stronger Bose-statistical coherence in the gas. When the applied field 
varies with a length larger than the value of $1/q_c(T)$, it is screened 
by the charged Bose gas with short-range repulsion. The increase of 
$q_c(T)$ in Eq.(28) means the growth of the  Bose-statistical coherence 
in cooling the system.

е


%
%


\begin{thebibliography}{99}
\bibitem{dia} Note that ODFRO is a different concept from the diagonal 
      long-range order (DLRO) found in various ordered state.
\bibitem{flu} The fluctuation occurring at the vicinity of $T_c$ in 
      the superconductors is smaller than that appearing in the liquid helium 4 
       close to $T_{BEC}$.
\bibitem{bog} It is well known that the charged Bogoliubov model, 
      in which $\Phi_{p=0}$ is replaced by the c-number in the 
      hamiltonian of Eq.(2), shows the Meissner effect like the 
      superconductors (Eq.(13)) at the first order of $g$.
       A theory without such a replacement is a purpose of this paper.
\end{thebibliography}
\end{document}